\documentclass[%
 reprint,
superscriptaddress,
nofootinbib,
 amsmath,amssymb,
 aps,
]{revtex4-1}
\usepackage{xcolor}
\usepackage{amsmath,amssymb}
\usepackage{graphicx}
\usepackage{dcolumn}
\usepackage{bm}
\usepackage{hyperref}

\begin{document}

\title{Observations of compact stars and fermion-boson stars with a quartic self-interaction}

\author{Susana Valdez-Alvarado} 
 \email{svaldeza@uaemex.mx}
\affiliation{%
Facultad de Ciencias de la Universidad Aut\'onoma del Estado de M\'exico  (UAEM\'ex.), Instituto Literario No. 100, C.P. 50000, Toluca, Estado de M\'exico, M\'exico}

\author{Alejandro Cruz-Osorio} 
 \email{aosorio@astro.unam.mx}
\affiliation{Instituto de Astronom\'{\i}a, Universidad Nacional Aut\'onoma de M\'exico, AP 70-264, Ciudad de M\'exico 04510, M\'exico}

\author{J. M. D\'avila} 
 \email{jmdavilad@uaemex.mx}
\affiliation{%
Facultad de Ciencias de la Universidad Aut\'onoma del Estado de M\'exico  (UAEM\'ex.), Instituto Literario No. 100, C.P. 50000, Toluca, Estado de M\'exico, M\'exico}

\author{L. Arturo Ure\~{n}a-L\'{o}pez} 
 \email{lurena@ugto.mx}
\affiliation{%
Departamento de F\'isica, DCI, Campus Le\'on, Universidad de
Guanajuato, 37150, Le\'on, Guanajuato, M\'exico.}

\author{Ricardo Becerril}%
 \email{ricardo.becerril@umich.mx}
 \affiliation{Instituto de F\'isica y Matem\'aticas, Universidad Michoacana de San Nicol\'as de Hidalgo. Edif. C-3, 58040 Morelia, Michoac\'an, M\'exico}

\date{\today}

\begin{abstract}
    We investigated the possibility that compact stars could be described by a fermion-boson star with a quartic self-interaction in the boson sector. Specifically, by varying the polytropic constant $K$ and adiabatic index $\Gamma$ in the polytropic equation of state, the boson mass $\mu$, and the self-interaction parameter $\Lambda$, we construct equilibrium configurations of these mixed-stars with total mass compatible with the mass constraints obtained from observational data of the collaborations NICE, NICER/XMN-Newton, and LIGO. Our work confirms that the addition of a boson sector eases the comparison of neutron star models with gravitational events related to compact objects and that in such a case observations may have preference for a positive self-interaction in the boson sector.
\end{abstract}

\maketitle

\section{Introduction \label{sec:introduction}}

Objects composed of a combination of fermions and bosons were first studied by Henriques and Liddle in~\cite{Henriques:1989ar,Henriques:1990xg,Henriques:1989ez} (see also~\cite{Henriques:2003yr,Lopes:1992aw}), and their properties have been further explored in recent work~\cite{ValdezAlvarado:2012xc,Valdez-Alvarado:2020vqa,DiGiovanni:2020frc}. These objects are known as fermion-boson stars (FBS) and are solutions of the Einstein-Klein-Gordon-hydrodynamic (EKGH) system of equations\cite{2017LRR....20....5L}. 

The boson sector of FBS consists of a massive scalar field that is minimally coupled to gravity, with a boson mass $\mu$ and a self-interaction parameter $\Lambda$ in the scalar potential~\cite{Colpi:1986ye,1998PhRvD..58j4004B, Guzman:2004wj,2003CQGra..20R.301S}. The fermion sector is usually modeled with a polytropic equation of state (EOS) of the form $P(\rho) = K\rho^{\Gamma}$~\cite{Valdez-Alvarado:2020vqa}, where $K$ and $\Gamma$ take values that correspond to masses and compactness in the range of neutron stars (NS). The stability of FBS has been verified for a wide range of parameters, allowing for a variety of masses and sizes in the resulting objects~\cite{Valdez-Alvarado:2020vqa,DiGiovanni:2020frc}.

On the other hand, the successful detection of gravitational waves by the LIGO and Virgo scientific collaborations (LVC) has significantly widened the window to observe our universe and to understand gravity itself, but some of the gravitational wave events have brought new theoretical challenges and opportunities. The GW190814 event ~\cite{LIGOGW1908142020}, for example, involved a black hole (BH) with a mass of $22.2 M_{\odot} - 24.3 M_{\odot}$ and a spin of $\le 0.07$, accompanied by a compact object with a mass of $2.5 M_{\odot} - 2.67 M_{\odot}$. This binary source of a gravitational wave has the smallest mass ratio ever measured for a similar system, at $0.112^{+0.008}_{-0.009}$. The nature of the second object is unknown, and there are no constraints on its radius, so it could be either a very light black hole or a very heavy neutron star (NS), which would make it the heaviest NS ever observed.

The peculiarities of events GW170817 and GW190814 have been used to set limits on the maximum mass and radii of generic neutron stars~\cite{Rezzolla2018,Nathanail2021,Most:2018hfd}, which in turn have been used to constrain the parameters of polytropic fluids in models~\cite{Arroyo-Chavez:2020oya}. More recent constraints on the properties of neutron stars have been found from observations of the NICER telescope~\cite{Riley2021, Raaijmakers2021,Miller2019}. In~\cite{2022PhRvD.105f3005D}, it was already suggested that an FBS, with a boson mass of the order of $\mu \sim 10^{-9} \mathrm{eV}$, could help explain the constitution of the self-gravitating objects behind the occurrence of these events, without self-interaction in the boson sector (i.e. $\Lambda =0$).

In this paper, we investigate the possibility that the secondary component of the event GW190814 is a FBS with a quartic self-interaction in the boson sector. For this, we have kept in mind four facts: (a) the mass of the secondary object for the LIGO gravitational wave signal GW190814 lies in the interval $[2.5,2.67]~ M_{\odot}$; (b) for Rezzolla – Nathanail simulations for neutron stars, its mass is located in the interval $[2.0,2.3]~M_{\odot}$; (c) for the NICER collaborations its corresponding mass interval is $[1.95,2.2]~M_{\odot}$ for \texttt{PSRJ0030+0451} and $[1.2,1.65]~M_{\odot}$ for \texttt{PSJ0740+6620}; (d) ultra-light bosons have been proposed as candidates to be dark matter in our universe~\cite{Matos:1999et,Matos:2000ng,PhysRevLett.85.1158,PhysRevD.64.123528,PhysRevD.53.2236,Matos:2004rs,Matos:2008ag,:2012xeMagana,Hui:2016ltb,2019FrASS...6...47U,2021ARA&A..59..247H,matos2023short}, and then these new events could give hints about their possible existence. We constructed mixed star configurations selecting $\Gamma$ and $K$ in such a way that our numerical results yield total masses that match the bounds mentioned in (a), (b), and (c) above. Furthermore, we also consider the constraint of the sound speed $c_s<1$ (in natural units) in the fluid sector of the FBS.

This paper is structured as follows. Section~\ref{sec:equations} outlines the equations of motion of the coupled Einstein-Klein-Gordon-hydrodynamic system and the boundary conditions that allow us to construct equilibrium configurations of FBS with a given quartic self-interaction in the boson sector. We also discuss the general numerical results in the case of FBS with total masses and radii similar to those of NS and the influence of the boson self-interaction in those quantities. 

In Sec.~\ref{sec:results} we present the comparison of our models with observational data of selected gravitational events, and discuss the general properties of the boson sector that can help the fluid star to be in more agreement with the data. Finally, in Sec.~\ref{sec:conclusions}, we provide a summary of our results and final remarks about our study.

\section{Equilibrium configurations 
\label{sec:equations}}

We model the boson sector of the FBS of interest with a complex scalar field $\phi$ that has a scalar potential $V( \phi)$. The fermion sector is represented by a perfect fluid with rest-mass density $\rho$, pressure $P$, internal energy $\epsilon$ and $4$-velocity $u^{\mu}$. The (relativistic) equations of motion for this system, expressed in geometrical units with $G=c=\hbar=1$, are as follows:
\begin{subequations}
\label{eq:ekg}
\begin{eqnarray}
 G_{\mu\nu} = 8\pi \left(T^{(\phi)}_{\mu\nu} + T^{(f)}_{\mu\nu} \right) \, , \quad \Box^2\phi - \phi\frac{dV(\phi)}{d(|\phi|^2)}=0 \, , \label{eq:ekg-a} \\
\nabla_\mu T^{(f) \mu\nu}=0 \, , \quad \nabla_{\mu}(\rho
u^{\mu})=0 \, , 
\label{eq:ekg-b}
\end{eqnarray}
where the stress-energy tensors of the bosonic and fermionic components are, respectively,
\begin{eqnarray}
    T_{\mu \nu}^{\phi} &=& \frac{1}{2} \left( 
\partial_{\mu}\phi\partial_{\nu}\phi^* \right) -\frac{1}{2} g_{\mu\nu}
\left( \partial^{\alpha}\phi^*\partial_{\alpha}\phi+2V \right) \, , \\
    T_{\mu \nu}^f &=& \left[\rho(1+\epsilon)+P \right]u_{\mu}u_{\nu}+Pg_{\mu\nu}\, .
\end{eqnarray}
\end{subequations}
The boson sector is endowed with a quartic scalar potential of the form
\begin{equation}
V(\phi) = \frac{\mu^2}{2} |\phi|^2 + \frac{\lambda}{4} |\phi|^4 \, ,
\label{eq:potential}
\end{equation}
that represents an ensemble of boson particles with mass $\mu$ and a self-interaction parameter $\lambda$, which may be either positive or negative. 

We are interested in equilibrium configurations, so we assume a static and spherically symmetric metric, with the line element $ds^2 = -\alpha^2(r)dt^2 + a^2(r)dr^2 + r^2 d\Omega^2$. The scalar field is expressed in the standard harmonic form $\phi(t,r)=\phi(r)e^{-i\omega t}$, with $\omega$ being the characteristic frequency of the solution, while all fermionic variables depend only on the radial coordinate $r$. To simplify the equations, we introduce a set of new variables: $\Omega=\omega/\mu$, $\sqrt{4\pi} \phi \to \phi$, $\Lambda = \lambda/(4\pi \mu^2)$, $4\pi \rho \to \rho$ and $4\pi P \to P$, with which Eqs.~(\ref{eq:ekg}) explicitly become,

\begin{subequations}
\label{eq:ekg2}
\begin{widetext}
\begin{eqnarray}
a^\prime &=& \frac{a}{2} \left(\frac{1-a^2}{r} + a^2r\left[\left(\frac{\Omega^2}{\alpha^2} + 1 + \frac{\Lambda}{2}\phi^2 \right) \mu^2 \phi^2 + \frac{\Phi^2}{a^2} + 2\rho(1+\epsilon)\right] \right) \, , 
  \label{aprima}\\
\alpha^\prime &=& \frac{\alpha}{2}\left(\frac{a^2-1}{r} +
  a^2r\left[\left(\frac{\Omega^2}{\alpha^2} - 1
  -\frac{\Lambda}{2}\phi^2 \right) \mu^2 \phi^2 + \frac{\Phi^2}{a^2} +
  2P\right] \right),
  \label{alphaprima}
\end{eqnarray}
\end{widetext}
for the metric variables $\alpha$ and $a$, whereas for the field and fluid variables we get
\begin{eqnarray}
\phi^\prime &=& \Phi, 
\label{phiprima}\\
\Phi^\prime &=& \left(1 - \frac{\Omega^2}{\alpha^2} + \Lambda \phi^2 \right) a^2 \mu^2 \phi -
  \left(\frac{2}{r} + \frac{\alpha^\prime}{\alpha} -
                     \frac{a^\prime}{a}\right) \Phi, ~~~~
                     \label{BigPhiprima} \\
P^\prime &=& -\frac{\alpha^\prime}{\alpha} [\rho(1+\epsilon) + P] \, . 
\label{Pprima}
\end{eqnarray}
\end{subequations}
Here, a prime denotes derivatives with respect to the radial coordinate $r$ and the internal energy is defined as $\epsilon = P/(\rho(\Gamma -1)) $. 

The corresponding boundary conditions that guaranty regularity at the origin and asymptotic flatness at infinity are
\begin{subequations}
\label{eq:boundary}
\begin{eqnarray}
a(0) &=& 1\, , \quad \alpha(0) =1 \, \quad \lim \limits_{r \rightarrow \infty} a(r)=1 \, , \label{eq:boundary-a} \\
\phi(0) &=& \phi_0 \, , \quad \Phi(0)=0 \, , \quad \lim \limits_{r \rightarrow \infty} \phi(r) = 0\, , \label{eq:boundary-c} \\
\rho(0) &=& \rho_0 \, , \quad P(0) = K \rho^{\Gamma}_0 \, , \quad \lim \limits_{r \rightarrow \infty} P(r)=0 \, . \label{eq:boundary-d}
\end{eqnarray}
\end{subequations}
The total mass of the equilibrium configurations, $M_T$ is obtained using the Schwarzschild definition,
\begin{equation}
M_T =  \lim \limits_{r \rightarrow \infty} \frac{r}{2} \left[ 1 -\frac{1}{a^2(r)} \right] \, .
\label{Masa}
\end{equation}

The numerical results from all the equations are dimensionless and given in code units. We use the standard convention for NS studies, where $M_T=1$ is equal to one solar mass $M_\odot$ in physical units. To obtain the total physical mass $M$ and the radius $99\%$ $R_{99}$ of the equilibrium configurations, we use the following expressions,
\label{eq:physical}
\begin{equation}
    M = 1 \, M_\odot \, M_T \, ,  \quad R_{99} \simeq 1.47 \, \mathrm{km} \, r_{99} \, , \label{eq:physical-a}
\end{equation}
where $r_{99}$ is the value of $r$ containing $99\%$ of the total mass $M_T$ but given in code units. A typical solution of a NS is then obtained without a scalar field ($\phi_0 =0$) and for appropriate values of fluid quantities $\rho_0$, $K$ and $\Gamma$.

For the mixed case of a FBS, we only need to solve the equations of motion~\eqref{eq:ekg2} for different values of the central field density $\phi_0$, the boson mass $\mu$ and the self-interacting parameter $\Lambda$. In particular, the boson mass $\mu$ in geometrical units is related to its physical value $\mu_B$ by
\begin{equation}
    \mu = 0.75 (\mu_B c^2/10^{-10} \mathrm{eV}) \, . \label{eq:boson_mass}
\end{equation}

We solved the equations of motion of FBS for three pairs of values $\Gamma$ and $K$ namely \{($\Gamma$, $K$)\}= \{($2.8$,\, $5.6\times 10^4$), ($2.85$,\, $7\times 10^4$), ($2.9$,\, $9\times 10^4$)\}; for each single pair we considering the following intervals for their free parameters: $0\leq \phi_0 \leq 0.2$, $0.006\leq \rho_0 \leq 0.06$ and $-50 \leq \Lambda \leq 50$ ~\footnote{In physical units, the central density is of the order $1.1 \lesssim \rho_0/\rho_\mathrm{nuc} \lesssim 11$, where the nuclear density is $\rho_\mathrm{nuc} = 2.66 \times 10^{14} \, \mathrm{g}/\mathrm{cm}^3$ ($\rho_\mathrm{nuc} = 4.3 \times 10^{-4}$ in geometrical units).}. 

\begin{figure}
    \centering
    \includegraphics[width=0.49\textwidth]{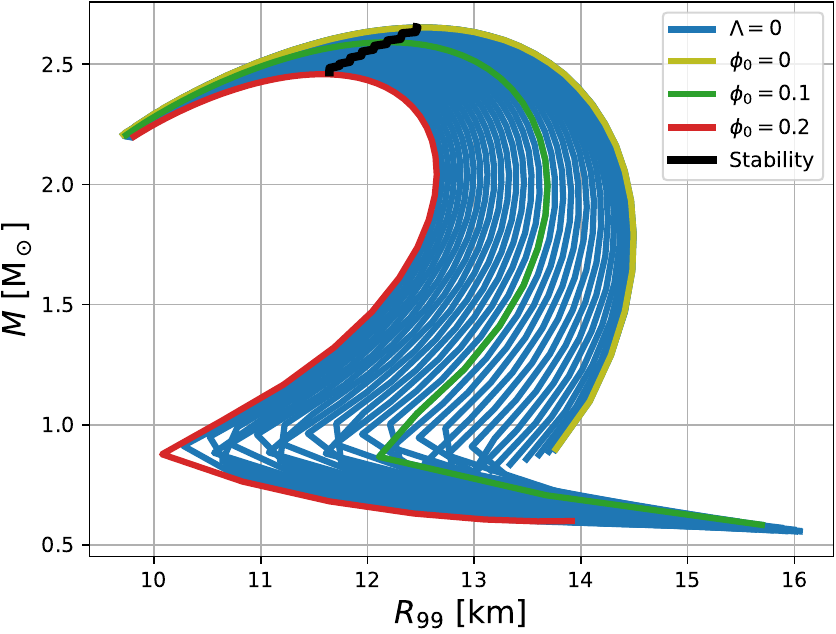}
    \includegraphics[width=0.49\textwidth]{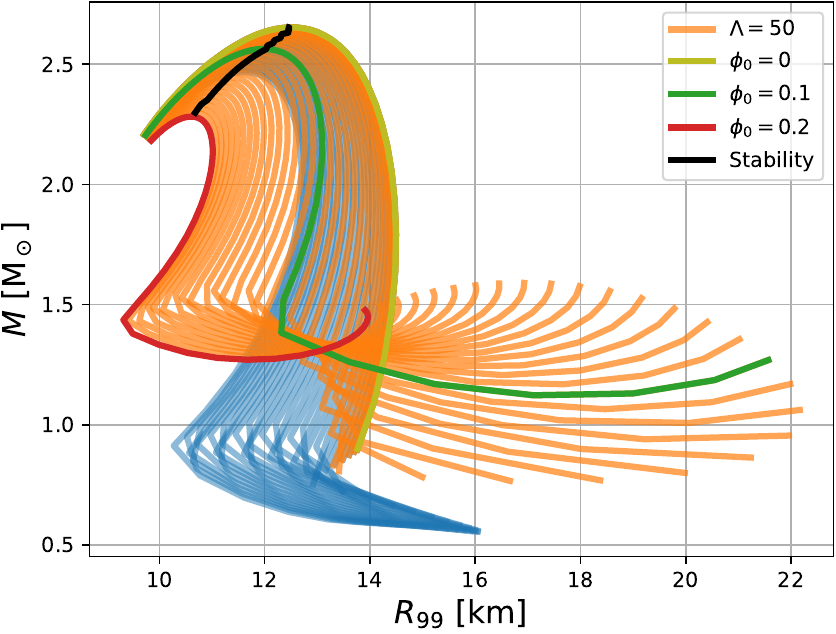}
    \includegraphics[width=0.49\textwidth]{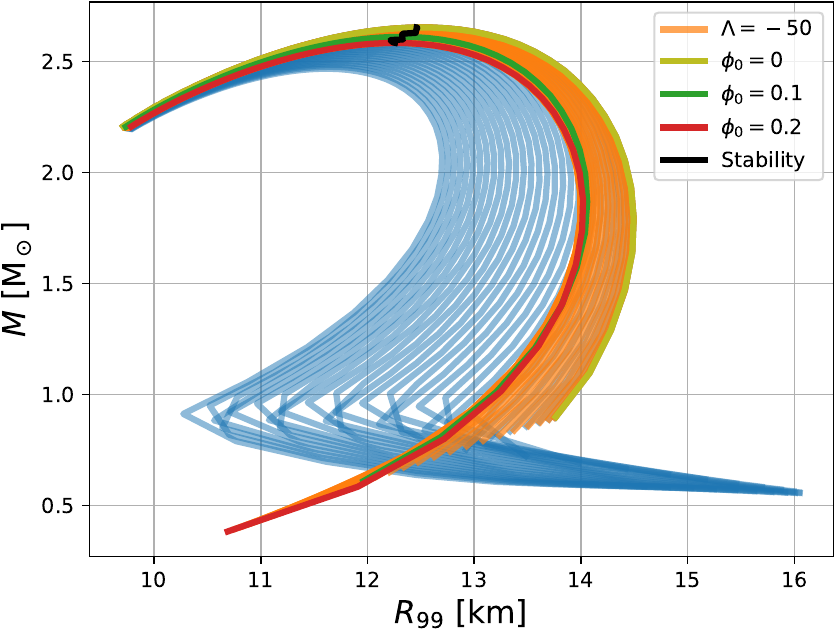}
    \caption{The top, middle and bottom panels of the figure show the equilibrium configurations of FBS for different values of the self-interaction parameter $\Lambda=0,50,-50$, respectively, in terms of their physical total mass $M$ and $99\%$ radius $R_{99}$. The cases $\Lambda=50,-50$ are superimposed on the case $\Lambda=0$ to compare their coverage of the parameter space. For reference, the configurations corresponding to the special values for the variation of the central field $\phi_0 =0,\,0.1,\,0.2$ are highlighted in each plot, while the black line represents the stability boundary. Further details can be found in the text.}
    \label{fig:FBC}
\end{figure}

The general behavior of the resulting configurations, in terms of their rescaled physical total mass $M$ and the $99\%$ radius $R_{99}$, is shown in Fig.~\ref{fig:FBC} for the cases $\Lambda=0,50,-50$ with $\Gamma=2.8$ and $K=5.6\times 10^4$. In all cases, the mass of the boson is $\mu=1$, which according to Eq.~\eqref{eq:boson_mass} means that $\mu_B c^2 \simeq 1.33 \times 10^{-10} \mathrm{eV}$. This is the expected mass for boson stars with a total mass of the order of solar masses and a radius in the range of tens of kilometers~\cite{Liebling_2023}. 

Each of the curves in the graphs represents a family of equilibrium configurations with a fixed value of $\phi_0$, and for which the central mass density $\rho_0$ is changed from bottom to top in each curve. Furthermore, as indicated in the graphs, the total area covered by the set of curves is limited by the extreme values of the central field $\phi_0$, whose corresponding curves have been highlighted. It should be noted that the curve $\phi_0 =0$ is purely fermionic and represents the standard NS for the chosen fluid parameters.

One last important feature is the stability of the equilibrium configurations. As previously studied in~\cite{ValdezAlvarado:2012xc}, see also~\cite{DiGiovanni:2020frc}, there is a standard method for determining the stability of FBS in terms of their number of bosons and fermions, which is slow and long. However, given that our FBS are fermion-dominated, we can take a shortcut by following the standard method for fermionic (fluid) stars. 

For a given $\phi_0$, we increase the central density $\rho_0$ until the maximum mass of the star is reached. All configurations with a central density higher than this should be unstable. It has been demonstrated in~\cite{Valdez-Alvarado:2020vqa} that this shortened method is consistent with the standard one for the small values of $\phi_0$ discussed here. For the sake of completeness, the configurations at the stability limit are also shown in Fig.~\ref{fig:FBC}.

When $\Lambda = 0$, the total mass and radius of stable FBS are in the ranges of $0.73 < M/M_\odot < 3$ and $10 < R_{99}/\mathrm{km} < 22$, respectively. The presence of the self-interaction parameter increases the area of the region covered by the curves for $\Lambda=50$ and decreases it for $\Lambda=-50$, while the mass and radius ranges remain similar for the three cases shown in Fig.~\ref{fig:FBC}. There is some overlap of the regions in the different cases, which implies that the self-interaction parameter has some degree of degeneracy with the central density for certain values of the mass and radius of the configurations.

For a positive self-interaction parameter, the mass of a boson star $M_B$ is given by $M_B \lesssim 0.165 \Lambda^{1/2} M_\odot/\mu$\cite{Colpi:1986ye} in the so-called Thomas-Fermi limit $\Lambda \gg 1$, and then $M_B \lesssim 8 M_\odot$ for $\Lambda=50$. This shows that the boson part of the FBS can contribute more to the mass of the star for the same value of $\phi_0$, and then the bundle of curves in the middle panel of Fig.~\ref{fig:FBC} can cover a larger portion of the parameter space. 

On the contrary, for a negative self-interaction parameter the upper bound on the total mass in the Thomas-Fermi limit $|\Lambda| \gg 1$ is $M_B < 1.4 |\Lambda|^{-1/2} M_\odot /\mu$~\cite{2011PhRvD..84d3531C}, so that $M_B < 0.2 M_\odot$ for $\Lambda=-50$. Being the mass contribution of the boson star this small, a narrow bundle of curves close to the anchor fermion star is to be expected, as the one shown in the bottom panel of Fig.~\ref{fig:FBC}.

\section{Comparison with observations \label{sec:results}}
Our goal in this section is to find regions in this parameter space for which the values of the total mass $M$ and radius $R_{99}$, of the resultant configurations, are compatible with those of NS as inferred from theoretical models and the NICER's observations.

\begin{figure*}
    \centering
    \includegraphics[width=\textwidth]{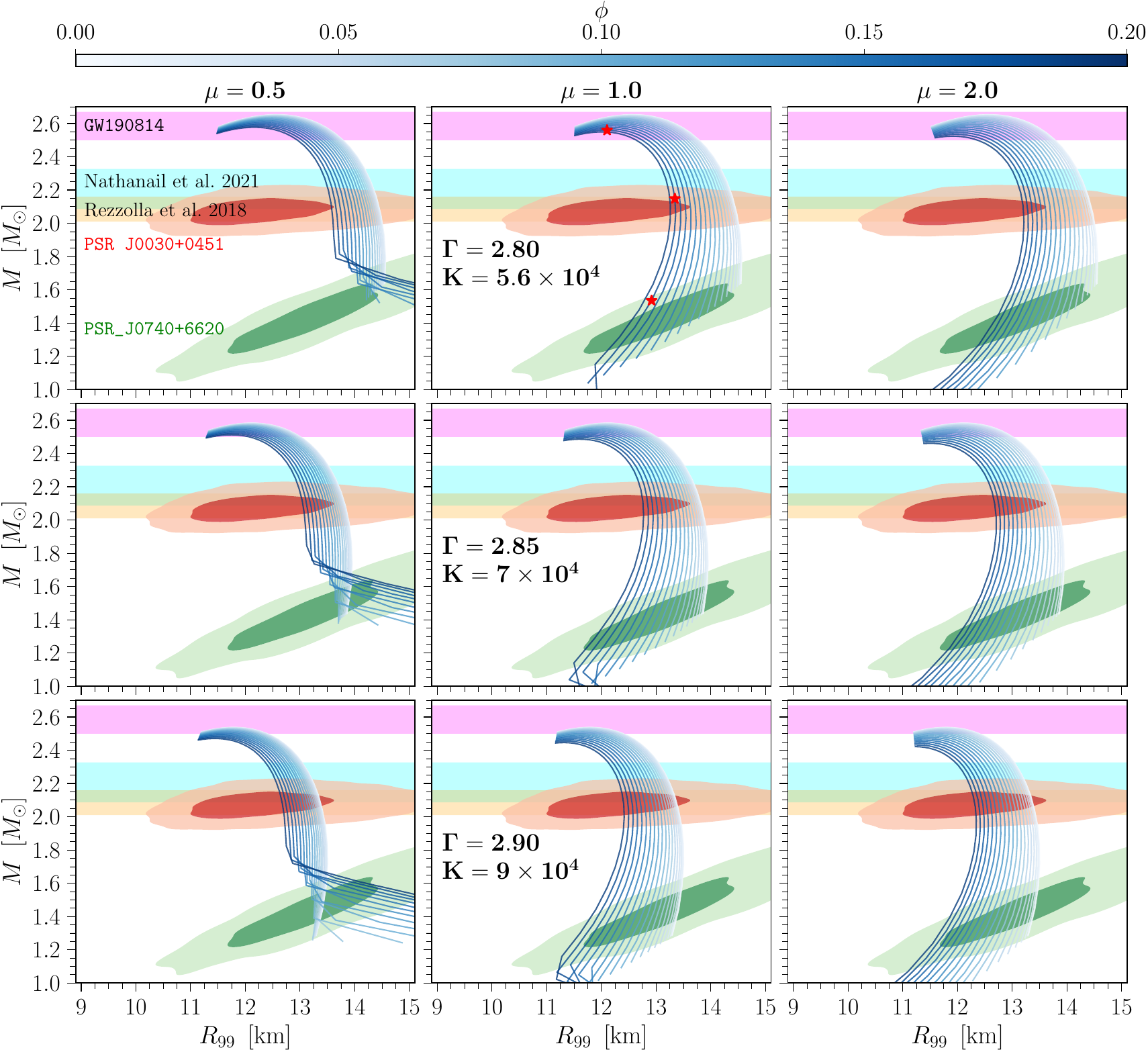}
    \caption{We show a comparison between plots of mixed stars masses vs radii (blue-white curves) with $\Lambda=10$ and the observational data obtained from NICER  (red and green regions) and LIGO (pink horizontal band) collaborations as well as from simulations performed by Rezzolla (green-yellow band) and Nathanail (cyan band). Each curve was constructed numerically taking a fixed value for $\phi_0$ and varying the central density $\rho_0$ in the range $[0.006,0.06]$. Each row corresponds to a given set of fluid parameters $(\Gamma,N)$ as indicated on the labels, and each column gives the results for a fixed value of the boson mass $\mu$. See the text for more details.}
    \label{fig:MvsR}
\end{figure*}

\begin{figure*}
    \centering
    \includegraphics[width=\textwidth]{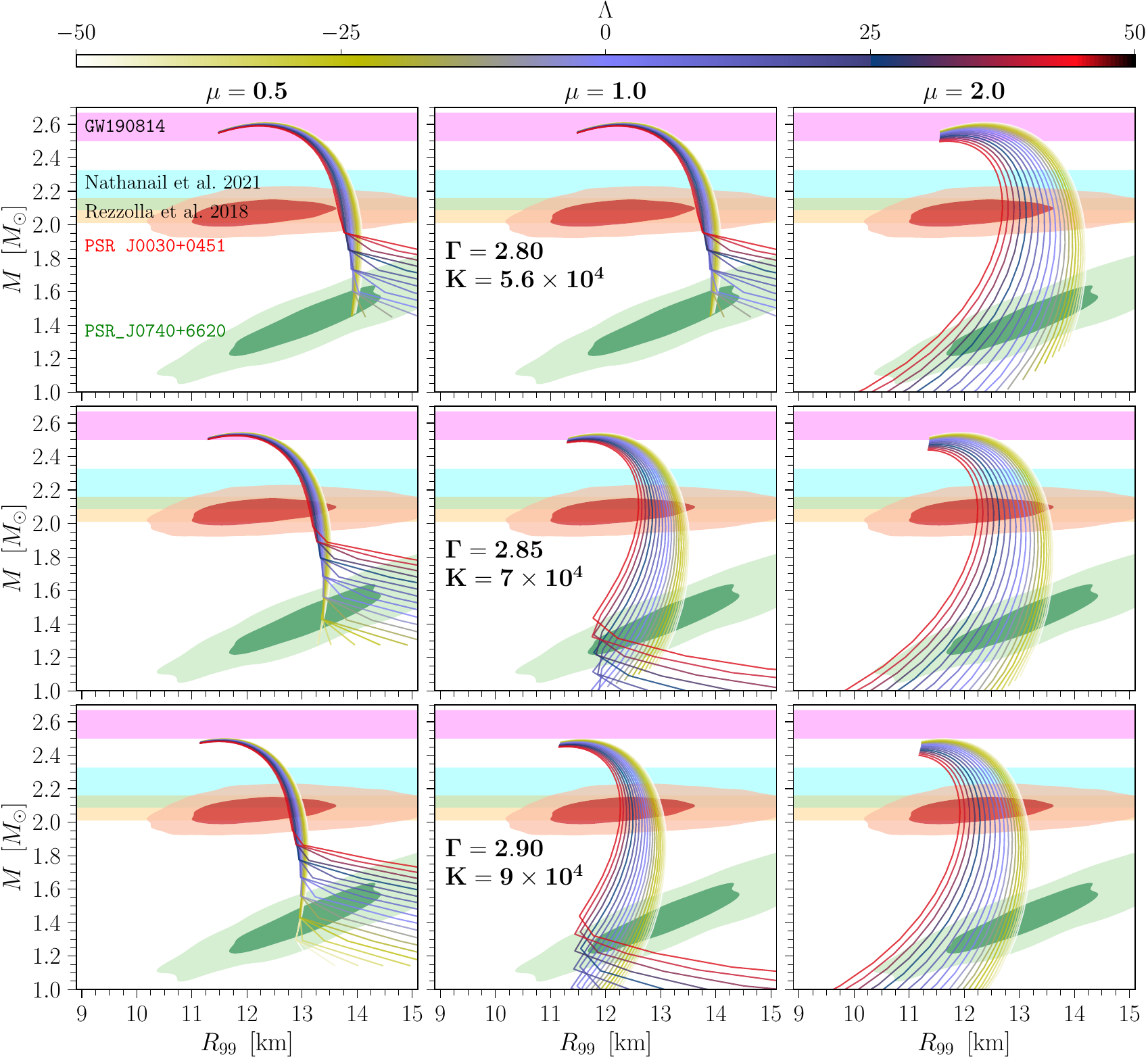}
    \caption{Comparison between families of mass-radius diagrams of mixed stars (multicolored curves) with $\phi_0=0.1$ and the observational data obtained from NICER  (red and green regions) and LIGO (pink horizontal band) collaborations as well as from simulations performed by Rezzolla (green-yellow band) and Nathanail (cyan band). Each curve was constructed numerically taking a fixed value for $\Lambda$ and varying $\rho_0$ in the range $[0.006,0.06]$. Each row corresponds to a given set of fluid parameters $(\Gamma,N)$ as indicated on the labels, and each column gives the results for a fixed value of the boson mass $\mu$. See the text for more details.}
    \label{fig:MvsRa}
\end{figure*}

Figures~\ref{fig:MvsR} and~\ref{fig:MvsRa} display the masses and radii of FBS obtained in physical units, superimposed on the properties of objects detected by different observational events. The green and red regions represent the density distributions determined by the NICER Collaboration for PSR J0740 + 6620~\cite{Riley2021,Raaijmakers2021} and PSRJ0030 + 0451~\cite{Miller2019}, respectively. The yellow and cyan regions indicate the constraints on NS mass values obtained from simulations of a large number of polytropic models conducted by Rezzolla {\it et al.} \cite{Rezzolla2018} and Nathanail {\it et al.}~\cite{Nathanail2021} respectively. Whereas the green  band represents the overlap  region between both simulations. The pink region corresponds to the mass associated with the second compact object detected in GW190814~\cite{LIGOGW1908142020} by LIGO, whose source is unknown. 

The case with fixed value $\Lambda=10$ is presented in Fig.~\ref{fig:MvsR} in terms of the central field strength $\phi_0$. In all instances, the value $\phi_0 =0$ corresponds to the NS (white curve), which we also referred to before as the anchor fermion star, constructed from three sets of values of the fluid parameters $(\Gamma,K)$: $( 2.80,\, 5.6\times 10^4)$ (top row), $( 2.85,\, 7\times 10^4)$ (middle row) and $(2.90,\, 9\times 10^4)$ (bottom row), as indicated in each graph. Likewise, three values of the boson mass were considered for the solution of the boson part of the star: $\mu = 0.5$ (left column), $\mu = 1.0$ (middle column) and $\mu = 2.0$ (right column).

As noted in~\cite{2022PhRvD.105f3005D}, as we increase the value of $\phi_0$, we can cover a larger area and agree with the regions suggested by the different observations. Notice that the NS profile mainly determines the mass range of the whole set of equilibrium configurations, while the scalar field helps to extend the range in the size of the FBS, as already pointed out in the discussion of Figs.~\ref{fig:FBC}. In particular, the variation on the field strength $\phi_0$ seems to give the required degree of freedom (width) to cover the confidence regions of PSRJ0030 + 0451 and PSRJ0740 + 6620. It can be seen that the best options appear to be those with $(\Gamma,K) =( 2.85, 7\times 10^4), (2.90,9\times 10^4)$ and with the boson mass $\mu$ of the order of unity ($\mu_B c^2 \simeq 10^{-10} \mathrm{eV}$).

In Fig.~\ref{fig:MvsRa}, we present the masses and radii of FBS obtained with the central value of the scalar field fixed at $\phi_0=0.1$, and for different values of the self-interaction parameter $\Lambda$ in the range $(-50,50)$. Given that the central scalar field is non-zero, the anchor configuration in these cases is not an NS but one FBS with $\Lambda=0$, and then configurations with $\Lambda > 0$ ($\Lambda < 0$) have smaller radii (larger radii) in comparison to the anchor FBS. 

In contrast to the cases in Fig.~\ref{fig:MvsR}, the best scenarios seem to be those in the column $\mu =2$, for the three sets of values of the fluid parameters. The values of $\Lambda$ provide the extra width in the parameter space to cover the density regions suggested by the observations, since $\Lambda$ can also influence the mass and radius of the configurations.

It is clear from the curves in Figs.~\ref{fig:MvsR} and~\ref{fig:MvsRa} that the masses and radii associated with the numerical configurations are capable of crossing all the areas suggested by the observations. To illustrate this in more detail, we have selected three cases with $\Lambda=10$, $\phi_0=0.1125$, $K=5.6\times 10^4$, $\Gamma = 2.8$, and $\mu=1$, for which the corresponding three FBS configurations have masses and radii that match the NICER and LIGO data. 

\begin{figure*}
    \centering
    \includegraphics[width=0.95\textwidth]{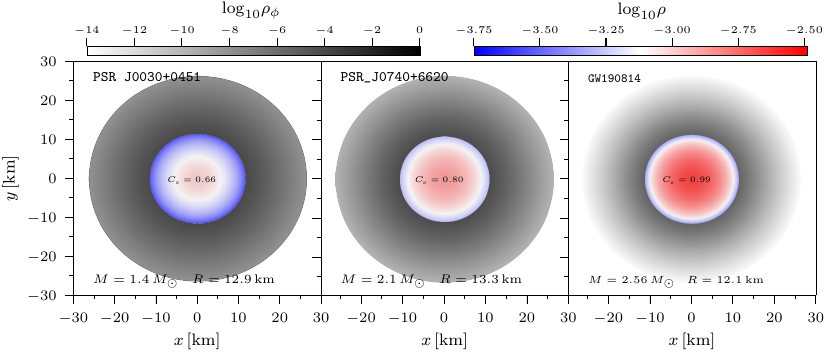}
    \caption{Two dimensional morphology of representative fermion-boson stars to model three observational constraints. The left panel shows an FBS with mass $M=1.4\, M_{\odot}$ and size $R_{99}=12.9\, km$, which meets the NICER observational constraints for the pulsar \texttt{PSRJ0030+0451}. The middle panel is an FBS with mass $M=2.1\ M_{\odot}$ and radius $R=13.3\ km$ to model the pulsar \texttt{PSRJ0740+6620}, satisfying the mass estimated by theory and NICER observations. The right panel shows the morphology of the FBS to model the event \texttt{GW190814} with mass $M=2.56\ M_{\odot}$ and radius $R_{99}=12.1\ km$. The fermion component is represented by the rest-mass density, and the boson component by the scalar field density, both in logarithmic scale. The three FBS are modeled by the same family of solutions (see the red stars in Fig.~\ref{fig:MvsR}), by fixing the self-interaction parameter $\Lambda=10$, the central strength of the scalar field $\phi_0=0.1125$, the equations for the state parameters $\Gamma=2.8$ and $K_{\rm new}=5.6\times 10^{4}$, and the mass of the boson $\mu=1$. See the text for more details.}
    \label{CINCO}
\end{figure*}

These cases are represented by the red stars in Fig.~\ref{fig:MvsR}, and their internal morphology is presented in Fig.~\ref{CINCO}. The gray color map represents the boson density, whereas the rest-mass density of fluid is represented in the blue-red scale. In general, we see that the contribution of the boson sector decreases as the configuration is more massive and compact, as a result of the concentration induced by the gravitational force of the fluid component, which is also the dominant one. The central sound speed of the FBS is also shown, which was calculated using $c^{2}_{\rm s}= P \gamma (\gamma-1)/[P\gamma - \rho (\gamma-1)]$. Even for the most extreme case with $M=2.56 \, M_\odot$, the sound speed remains below unity.

As mentioned above, the secondary component of the recently reported event GW190814 is of unknown nature, but its extreme mass indicates that it could be a light BH or a very heavy NS. However, our numerical results in Figs.~\ref{fig:MvsR} and~\ref{fig:MvsRa} make it plausible to identify this secondary component with an FBS with a quartic self-interaction.
  
\begin{figure*}
    \centering
    \includegraphics[width=0.48\textwidth]{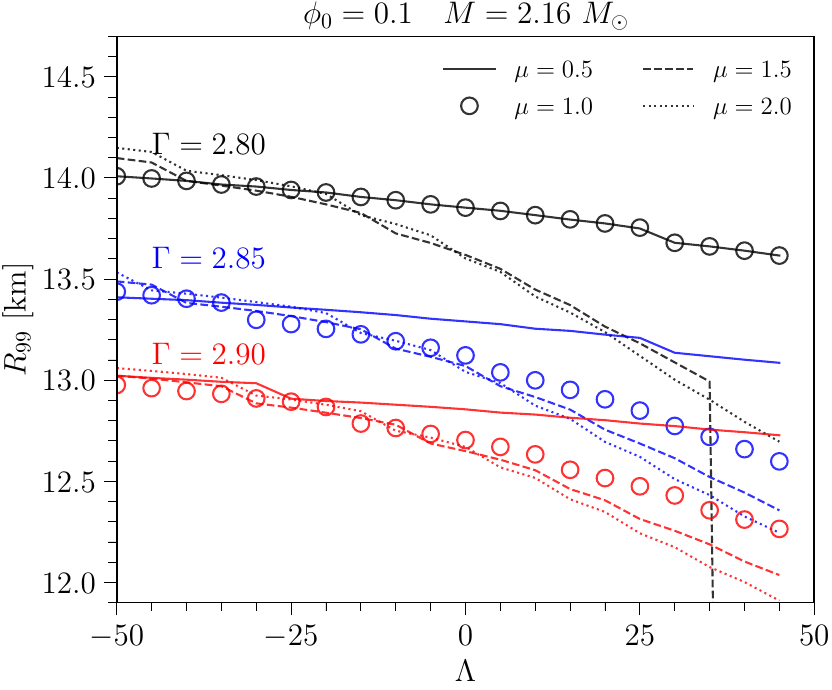}
    \includegraphics[width=0.48\textwidth]{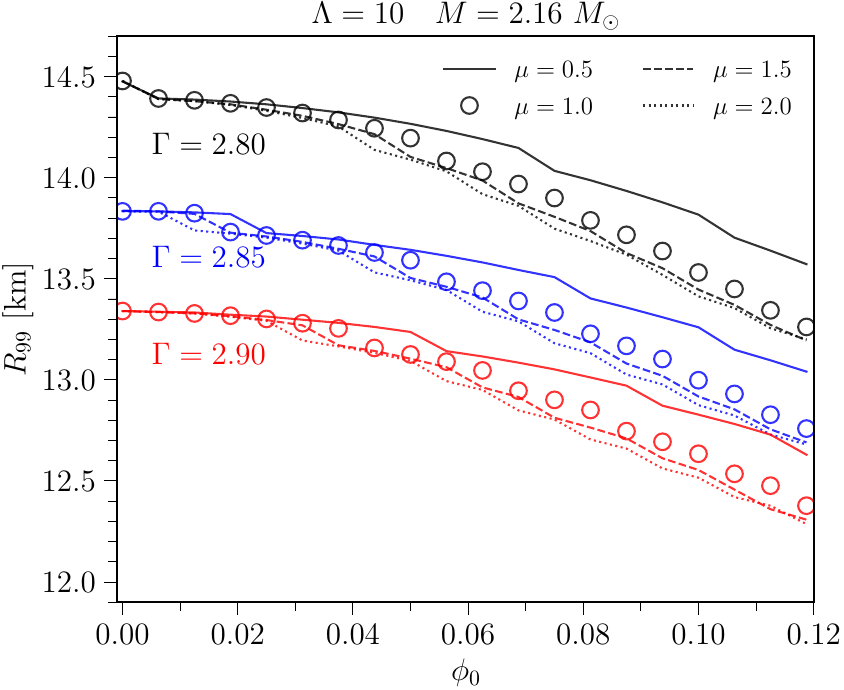}
    \caption{The size of the FBS as function of the self-interaction parameter (top panel) and the scalar field strength (bottom panel) for star with total mass $M=2.16\, M_{\odot}$. Changes in EoS are represented by black, blue, and red colors, while line styles show the effects of varying the boson mass $\mu$.  See the text for more details.
    }
    \label{fig:size}
\end{figure*}

To verify this more carefully, we have performed a detailed survey in the parameter space $\{\Gamma,K,\mu,\Lambda\}$ to determine the regions corresponding to an FBS with a fixed total mass of the order of $M \simeq 2.16 M_\odot$. The results are shown in Fig.~\ref{fig:size}, for the total radius $R$ as a function of the self-interaction parameter $\Lambda$ (top panel) and as a function of the central field density $\phi_0$ (bottom panel). The values of the polytropic parameters are the same as those considered in Figs.~\ref{fig:MvsR} and~\ref{fig:MvsRa} above.

In general, we note that the presence of the scalar field $\phi$, in terms of its central density or its self-interaction, reduces the size of the FBS, with the total radius reaching values on the order of $R_{99} \simeq 12 \, \mathrm{km}$. The most extreme cases are reached for the case $\mu =2$, together with the highest values of $\phi_0$ and $\Lambda$. For these extreme cases shown in Fig.~\ref{fig:size}, the compactness of the FBS is of the order of $\mathcal{C} =0.22$, which is approximately $16\%$ higher than the anchor NS alone.

\begin{figure*}
    \centering
    \includegraphics[width=0.48\textwidth]{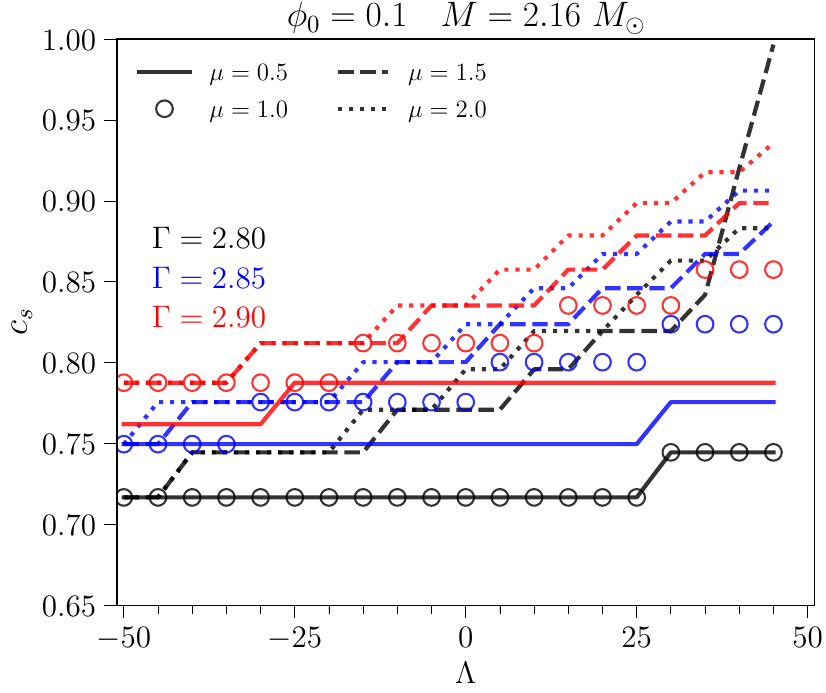}
    \includegraphics[width=0.48\textwidth]{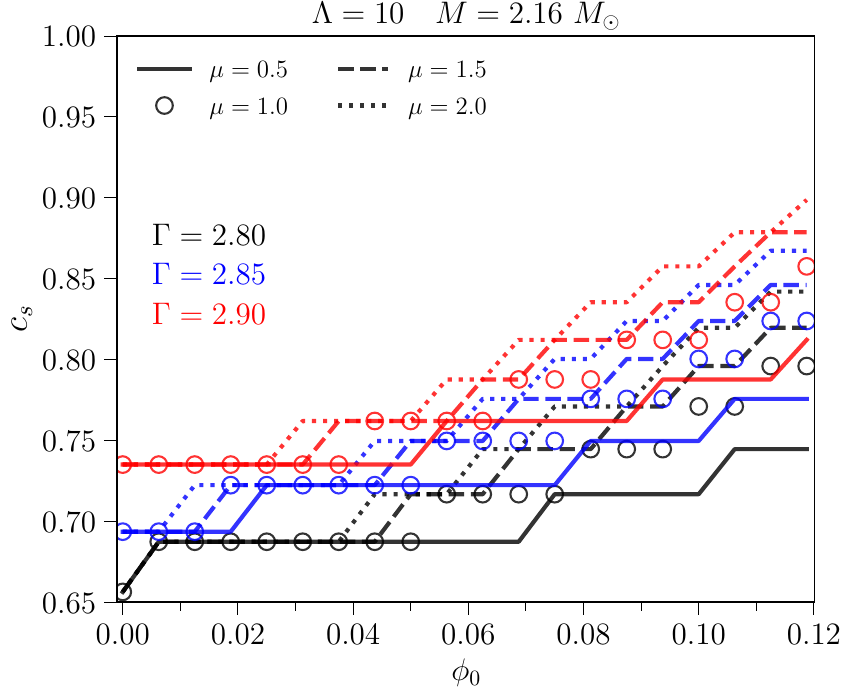}
    \caption{Sound speed at the center of the FBS as function of the self-interaction parameter (top panel), and as function of the amplitude at the star center. In all models, we fixed the mass of the FBS to $M=2.16\, M_{\odot}$, the corresponding sizes are reported in Fig.~\ref{fig:size}. The colors of the lines represent the three equations of states ($\Gamma=2.8,\, K=5.6\times10^{4}$), ($\Gamma=2.85,\, K=7\times10^{4}$), and ($\Gamma=2.9,\, K=9\times10^{4}$), respectively. The line styles correspond to different choices for the boson mass, $\mu=0.5,\ 1.0,\ 1.5,\ 2.0$. See the text for more details.
    }
    \label{fig:cs}
\end{figure*}

Another quantity of interest is the central sound speed $c_s$ of the FBS with $M=2.16\, M_{\odot}$, which we show in Fig.~\ref{fig:cs}. Similarly to the case of the compactness of the star, we see that the presence of the scalar field $\phi$ and its self-interaction $\Lambda$ increases the value of $c_s$, although in all cases the top sound speed is of the order of $c_s \simeq 0.9$. 

\section{Conclusions and final remarks \label{sec:conclusions}}

We have constructed FBS configurations with a self-interaction term in the boson sector and considered different values of their physical parameters to match the data for NS obtained from NICER, Rezzolla, and Nathanail collaborations as well as from the LIGO gravitational wave signal GW190814. For the fluid parameters, we picked values that yielded NS masses in the vicinity of the values reported for different observed objects, and then varied the boson parameters, mass and self-interaction, to find their influence in the mass, size, and morphology of the corresponding FBS.

In general terms, the variation of the boson parameters allows our models to cover a wider area in the parameter space of the observed objects, which means that the addition of a boson sector to a given NS configuration could help explain the presence of different types of object. According to our results, the best options seem to have a boson mass of order $\mu \simeq 1$ in our units, which means that the total mass of the corresponding boson star is of the same order of magnitude as that of the selected NS (or anchor NS, as we call it in the main text). Similarly, there is a preference for positive values of the self-interaction parameter, as a negative value of it makes it difficult to reconcile our theoretical results with observations.

Our work confirms and extends the results presented in~\cite{2022PhRvD.105f3005D}, in that the addition of a boson sector eases the comparison of NS models with different observations. We have already mentioned the preference for the value $\mu \simeq 1$, which is not surprising given the typical masses and sizes of compact objects suggested by the observations we used. The interesting new result is that the self-interaction, the next-to-leading order parameter in the description of the boson sector, in particular for the scalar field potential, should be positive. 

Some other comments are in turn. Our anchor NS was modeled as a polytropic fluid, for which the parameter values were selected according to previous studies of NS~\cite{Arroyo-Chavez:2020oya}. However, our study could also consider other types of equation of state for the fermion sector, such as those chosen in~\cite{2022PhRvD.105f3005D}, for which we would expect the same qualitative results obtained so far: preference of data for $\mu \simeq 1$ and $\Lambda \gtrsim 0$.

Ultralight bosons have been proposed as candidates for dark matter in our universe, and a boson with mass $\mu_B c^2 \sim 10^{-10} \, \mathrm{eV}$ would certainly qualify as ultralight, although such a mass would be 10 orders of magnitude larger than the preferred astrophysical value of $\mu_B c^2 \sim 10^{-22} \, \mathrm{eV}$~\cite{Hui:2016ltb,matos2023short}. Our new result $\Lambda \gtrsim 0$, if taken at face value, would mean that there is an extra constraint on dark matter models with axion-like bosons (with negative self-interaction), at least for the cases for which $\mu_B c^2 \sim 10^{-10} \, \mathrm{eV}$, as there are differentiable consequences at astrophysical and cosmological scales for the formation of cosmic structure from the sign of the self-interaction in the potential~\cite{2021PhRvD.103h3509M,2021JCAP...01..051L,2017PhRvD..96f1301C,2023MNRAS.521.2608M}.

As already noted in~\cite{2022PhRvD.105f3005D}, for comparison with observations, we just took the confidence regions obtained by other groups that considered models quite different from ours to fit the data. In this respect, our constraints should be considered conservative and more like a proof-of-concept for FBS with a self-interaction. A separate study with a direct comparison with raw data from the different observations would be necessary to have a more accurate judgment of the appropriateness of FBS and the role of their self-interaction in each of the reported gravitational events. This is work in progress, which we expect to report elsewhere.

%
\begin{acknowledgements}
S.V.A and J.M.D. acknowledge support from Programa de Desarrollo Profesional Docente de la Secretaria de Educaci\'on P\'ublica (PRODEP-SEP), under No. UAEM-CA-14 project. R.B. acknowledges partial support from C.I.C-UMSNH. LAU-L. acknowledges partial support from the Programa para el Desarrollo Profesional Docente; Direcci\'on de Apoyo a la Investigaci\'on y al Posgrado, Universidad de Guanajuato; CONACyT M\'exico under Grants No. 286897, No. 297771, No. 304001; and the Instituto Avanzado de Cosmolog\'ia Collaboration.
\end{acknowledgements}

%
\bibliography{paper}

\end{document}